\documentclass[10pt,a4paper,twocolumn]{extarticle} 

\usepackage{graphicx}
\usepackage{dcolumn}
\usepackage{bm}
\usepackage{gensymb}
\usepackage{amsmath}
\usepackage[affil-it]{authblk}
\usepackage{abstract}
\usepackage[english]{babel} 
\usepackage{siunitx} 
\usepackage[hmarginratio=1:1,top=25mm,left=18mm,columnsep=20pt]{geometry} 
\usepackage[noadjust]{cite}

\title{\vspace{-1.1cm}\large\textbf{Verification of a dynamic scaling for the pair correlation function during the slow drainage of a porous medium}}

\author[1]{M. Moura\thanks{Corresponding author (marcel.moura@fys.uio.no)}}
\author[1]{K. J. M\aa{}l\o{}y}
\author[1]{E. G. Flekk\o{}y}
\author[1,2]{R. Toussaint}
\affil[1]{PoreLab, Department of Physics, University of Oslo, Norway}
\affil[2]{Universit\'e de Strasbourg, CNRS, IPGS UMR 7516, F-67000 Strasbourg, France}

\begin{document}
\twocolumn[
  \begin{@twocolumnfalse}
    \maketitle
  \vspace{-0.3cm}
\begingroup\par
\centering
\begin{minipage}{0.8\textwidth}
\hspace{0.5cm} We give experimental grounding for the remarkable observation made by Furuberg et al. in Ref.~\cite{furuberg1988} of an unusual dynamic scaling for the pair correlation function $N(r,t)$ during the slow drainage of a porous medium. The authors of that paper have used an invasion percolation algorithm to show numerically that the probability of invasion of a pore at a distance $r$ away and after a time $t$ from the invasion of another pore, scales as $N(r,t)\propto r^{-1}f\left(r^{D}/t\right)$, where $D$ is the fractal dimension of the invading cluster and the function $f(u)\propto u^{1.4}$, for $u \ll 1$ and $f(u)\propto u^{-0.6}$, for $u \gg 1$.  Our experimental setup allows us to have full access to the spatiotemporal evolution of the invasion, which was used to directly verify this scaling. Additionally, we have connected two important theoretical contributions from the literature to explain the functional dependency of $N(r,t)$ and the scaling exponent for the short-time regime ($t \ll r^{D}$). A new theoretical argument was developed to explain the long-time regime exponent ($t \gg r^{D}$).
\end{minipage}

\vspace{0.5cm}
\end{@twocolumnfalse}
\par \endgroup
]
\saythanks

The slow drainage of a porous medium is characterized by a rich intermittent dynamics of invasion bursts, typically occurring at several time and length scales \cite{haines1930,maloy1992,furuberg1996}. Similar intermittent activity is observed in a wide variety of physical, biological and social systems \cite{burlaga1991,ruzmaikin1995,sneppen1995,gomes1998,liu1999,sethna2001,maloy2006,grob2009,planet2009,salazar2010,santucci2011,tallakstad2013,stojanova2014,gonzlez2017}. The ubiquity of intermittent phenomena is an indication that its origin is not expected to depend on specific system-dependent details. It is generally associated to the competition between an adaptive external driving force and an internal random resistance against that force \cite{maloy1992,moebius2012}. Energy is slowly injected by the external force during stable periods which are abruptly interrupted by sudden dissipative events occurring at a much faster time scale \cite{pomeau1980,pomeau1980_2,hirsch1982}. In the case of two-phase flows in porous media, the external force could come from a syringe pump or an applied pressure difference across the sample, while the internal resistance is caused by the narrower or wider pore-throats that are invaded during the flow \cite{haines1930,maloy1992,furuberg1996}.

The balance between viscous, capillary and gravitational forces \cite{meheust2002,grunde2005,or2008,polak2011} generates several interesting invasion patterns in porous media flows, ranging from compact invasion \cite{cieplak1990,trojer2015,holtzman2015} to fractal structures \cite{maloy1985,lenormand1985,lenormand1989,lenormand1989.2}. Although the pattern morphology has been well studied, very few studies have focused on its dynamical features \cite{maloy1992,furuberg1996,moebius2012,berg2013,moebius2014.2}, due in part to the difficulty of simultaneously analyzing detailed invasion data in both temporal and spatial domains. Single pore invasion events in rocks have only recently been imaged in real-time via modern X-ray microtomography techniques \cite{berg2013,bultreys2015}, but extending those techniques to study pattern formations in rocks at the larger scales, while keeping single-pore resolution, remains a challenge.

The experimental constraints fueled the development of several numerical algorithms. Invasion percolation (IP) \cite{wilkinson1983,stauffer1994}, diffusion-limited aggregation (DLA) \cite{witten1981,paterson1984,maloy1985} and anti-DLA \cite{paterson1984} have been employed in the simulation of, respectively, slow drainage (capillary fingering), fast drainage (viscous fingering) and stable imbibition (compact growth) \cite{lenormand1988}. In the standard IP model the invasion happens one pore at a time, always at the widest available pore-throat (with the smallest capillary pressure threshold). Using an IP model, Furuberg et al. \cite{furuberg1988} have addressed the following question: given that a reference pore located at position $\bm{r_0}$ is invaded at a time $t_0$, what is the probability that a second pore located at position $\bm{r_1}$ is invaded at some later time $t_1$? After the vanishing of transitional effects, the probability should only be a function of the relative quantities $r=\left|\bm{r_1}-\bm{r_0}\right|$ and $t=t_1-t_0$. The authors have defined a pair correlation function $N(r,t)$, such that $N(r,t)drdt$ gives the answer to the question, i.e., it is the probability of invasion of a pore located between distances $r$ and $r+dr$ and at a time between $t$ and $t+dt$ with respect to the invasion of the reference pore.  By considering theoretical arguments related to the normalizaion of $N(r,t)$ and the connection between $N(r,t)$ and the pair connectedness function, the authors have suggested  the form

\begin{equation}
N(r,t)=r^{-1}f\left(\frac{r^{D}}{t}\right) \:,
\label{eq:hat1}
\end{equation}
where the dynamic exponent $D$ corresponds to the fractal dimension of the invaded front \cite{mandelbrot1982,feder1988}. The IP simulations employed in that study numerically confirmed the validity of Eq.~(\ref{eq:hat1}) and found out additionally that the function $f(u)$ presents the unusual scaling
\begin{equation}
f(u) \propto \begin{cases}
u^{1.4} &\text{if $u \ll 1$}\\
u^{-0.6} &\text{if $u \gg 1$}
\label{eq:hat2}
\end{cases} \:.
\end{equation}
In the current work, we have demonstrated these results experimentally, almost 30 years after the original findings of Furuberg et al. \cite{furuberg1988}.

\begin{figure}
\centering
	\includegraphics[width=\linewidth]{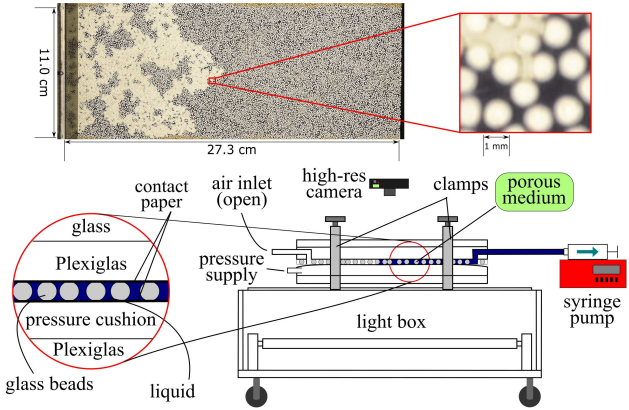}
	\caption{(color online) Diagram of the experimental setup. On top we show an image of the flow and in the detail a zoom in section of the porous network, the defending wetting phase (liquid, blue) and the invading non-wetting phase (air, white).}
	\label{fig:diagram}
\end{figure}

Fig.~\ref{fig:diagram} shows a diagram of the experimental setup. A quasi-2D porous network is formed by a monolayer of glass beads, with diameters $a$ in the range \mbox{$1.0\si{mm}<a<1.2\si{mm}$}, randomly placed in the gap of a modified Hele-Shaw cell. The beads are kept in place by a pressurized cushion on the bottom plate of the cell. The porous medium is initially saturated with a viscous liquid (wetting phase), composed of glycerol ($80\%$ in weight) and water ($20\%$ in weight), having kinematic viscosity $\nu = 4.25\cdot10^{-5}\, \si{m^2/s}$, density $\rho = 1.205\,\si{g/cm^3}$ and surface tension $\gamma = 0.064 \,\si{N/m}$ (see Ref.~\cite{moura2015} for additional details). The outlet channel is connected to a syringe pump, from which liquid can be slowly withdrawn at a constant rate of $q = 0.0050\,\si{ml/min} \approx 0.14\,\si{pore/s}$, thus assuring that the dynamics happens in the capillary regime, in which capillary forces dominate over viscous ones \cite{lenormand1989.2}. Air (non-wetting phase) enters from a width-spanning inlet channel, which is kept open to the atmosphere. Once the capillary pressure (difference in pressure between the non-wetting and wetting phases) is large enough to overcome the threshold associated with a given pore-throat, the invasion of one or more pores happens, and viscous pressure drops are triggered within the medium, thus dissipating energy \cite{maloy1992,grunde2011}. Pictures are taken every $34\,\si{s}$ which leads to an average number of $K\approx 4.7$ invaded pores per image (average pore invasion time of $t_p = 7.2\,\si{s}$). The capillary number is, $Ca=\rho\nu q/\Sigma \gamma = 6.1\cdot10^{-7}$, where $\Sigma=1.1\cdot10^{-4}\, \si{m^2}$ is the cross section of the model. The choice of a slow constant withdrawal rate reflects the fact that in the IP algorithm employed in Ref.~\cite{furuberg1988}, viscous pressure drops are neglected and exactly 1 pore (numerical site) is invaded per time step.

  \begin{figure}
      \centering
      \includegraphics[width=\linewidth]{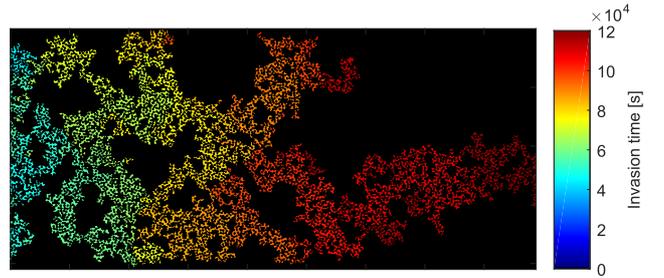}
      \caption{(color online) Spatiotemporal map of the invasion up to breakthrough (the average flow direction is from left to right). The colormap shows the elapsed time for the invasion of a given pore (in seconds). The experiment lasts $\approx33h$.}
      \label{fig:detail_invasion}
  \end{figure}

Fig.~\ref{fig:detail_invasion} shows the spatiotemporal map of the invasion. The flow is from left to right but statistically similar results should be found if the flow were in the opposite direction since the model is prepared randomly. The experiment stops at breakthrough, i.e., when the air phase first percolates (equivalent to the infinite cluster in IP). Only the central $90\%$ of the model's length (inlet--outlet direction) is considered, in order to avoid boundary effects which have been observed close to the inlet and outlet \cite{moura2015}. The information content in this map is similar to that in the IP simulations, namely the position and invasion times of all pores, thus allowing the computation of the pair correlation function $N(r,t)$. In order to precisely locate the pores, we performed a Delaunay triangulation \cite{lee1980} of the points marking the centers of all glass beads and then identified the centroid of each Delaunay triangle as a pore center. In the IP simulations \cite{furuberg1988}, the time $t$ is naturally measured as the number of invaded pores, whereas in our experiments it is given by the image number $t_I$. The conversion between $t_I$ and $t$ is $t_I=t/K$, where $K$ is the average number of pores invaded per image. This linear conversion works best for large times (several pore invasions) since the intermittency of the Haines jumps \cite{haines1930} is relevant for short times.

For a given reference pore, one can produce a histogram of the euclidean distances $r$ of all pores invaded between times $t_I$ and $t_I+\Delta t_I$ after its own invasion. By treating each invaded pore in the system as a reference for the production of one such histogram and adding them up, we obtain the function $N(r,t)$ (apart from a normalization factor). We have chosen $\Delta t_I=2$ images to guarantee that a small number of pore invasions (given on average by $\Delta t_I K \approx 9.4$)  would be present on each histogram. Different values of $\Delta t_I$ in the range $1\leq \Delta t_I\leq 5$ were tested and the results presented did not change significantly, since the experiment duration ($\approx 33\,\si{h}$) is much longer than the time between images ($\approx 34\,\si{s}$). It is worth mentioning that, due to the fast nature of the Haines jumps \cite{haines1930}, if one wants to capture the \emph{exact} invasion time of a pore, one would need to rely on high-speed imaging, which would then dramatically constrain the total duration of the experiment. Such level of accuracy was not needed in this study.
\begin{figure}
	\centering
	\includegraphics[width=\linewidth]{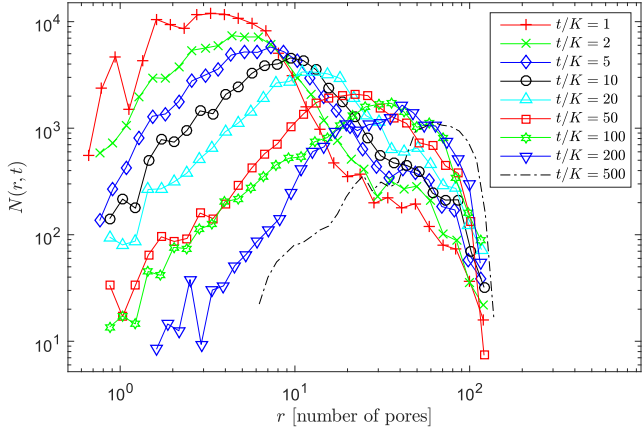}
	\caption{(color online) Pair correlation function $N(r,t)$  as a function of the distance $r$ for 9 times $t$ (shown in the legend).}
	\label{fig:hat1_2}
\end{figure}

\begin{figure}
	\centering
	\includegraphics[width=\linewidth]{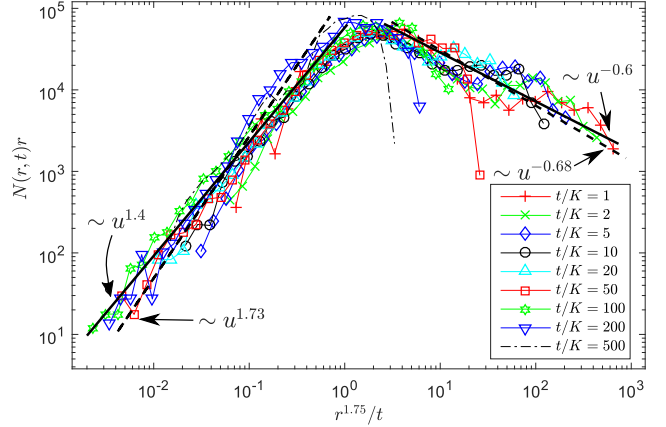}
    \caption{(color online) Data collapse for the product $N(r,t)r$ as a function of the reduced variable $u=r^{D}/t$ for 9 different fixed times (see legend). The scaling regimes for $u \ll 1$ and $u \gg 1$ as reported in Ref.~\cite{furuberg1988} (thick solid lines) and their theoretical predictions (thick dashed lines) are also shown.}
	\label{fig:hat2}
\end{figure}

The measured function $N(r,t)$ is shown in Fig.~\ref{fig:hat1_2} for $9$ fixed values of time $t/K=t_I$ (shown in the legend). The distance $r$ is measured in pore length units (euclidean distance divided by a characteristic interpore distance $r_p=0.92\,\si{mm}$). $N(r,t)$ presents a peak indicating a maximum probability of invasion at a certain relative distance $r_t$ which increases with time (since the air-liquid interface had more time to move, pores farther away can be invaded). The validity of Eq.~(\ref{eq:hat1}) is proved by using this equation to collapse the data from Fig.~\ref{fig:hat1_2}. The result, shown in Fig.~\ref{fig:hat2}, indicates that the product $N(r,t)\:r$ is indeed a function only of the reduced variable $u=r^D/t$ and not of $r$ and $t$ separately (the fractal dimension of the invasion cluster was measured to be $D=1.75$). This confirms the validity of the dynamic scaling in Eq.~(\ref{eq:hat1}). The fact that the product $N(r,t)\:r$ is peaked around $r^D/t=1$ (with $r$ and $t$ measured in terms of pore length and average pore invasion time units) indicates that the most probable place for invasion occurs at a distance $r=t^{1/D}$, as noted in \cite{furuberg1988}. If one chooses another set of measuring units, the maximum changes to $r_p^D/t_p$, where $r_p$ and $t_p$ are the typical interpore distance and pore invasion time in the new units. The scaling behavior of $N(r,t)$ could not be easily inferred from the separate curves in Fig.~\ref{fig:hat1_2}, due to the limited statistics, but becomes more visible after the data collapse in Fig.~\ref{fig:hat2}. Both scaling regimes, $u^{1.4}$ for $u \ll 1$ and $u^{-0.6}$ for $u \gg 1$, are well reproduced by the experiments (guide-to-eye thick solid lines). Indeed, even the deviation from the scaling observed as the dropping curves for large times and $u \gg 1$ (due to finite-size effects) are also in agreement with the simulations in \cite{furuberg1988}.

Next we analyze the origin of this scaling behavior. In the work of Roux and Guyon \cite{roux1989}, the distribution of temporal avalanches in the time series of threshold values from an IP model was used to propose an analytical prediction for the unusual scaling presented in Eqs.~(\ref{eq:hat1}) and (\ref{eq:hat2}). Later on, Maslov \cite{maslov1995} pointed out some inconsistencies in the assumptions used in Ref.~\cite{roux1989}, which interfered with their estimation for the exponents in Eq.~(\ref{eq:hat2}). Nevertheless, the reasoning in Ref.~\cite{roux1989} to justify the functional form behind Eqs.~(\ref{eq:hat1}) and (\ref{eq:hat2}) still applies. By taking into account the results of Maslov \cite{maslov1995}, into the analysis performed by Roux and Guyon \cite{roux1989}, we obtain a consistent value for the short-time exponent in Eq.~(\ref{eq:hat2}) (for $u\gg 1$ or $t \ll r^{D}$). The analysis for the long-time exponent (for $u\ll 1$ or $t \gg r^{D}$) will require some additional considerations, as shown later.

Following Ref.~\cite{roux1989}, for a given time interval $t$, the distribution of backwards temporal avalanches $\Theta$ (as defined in \cite{roux1989} and also in \cite{maslov1995}) that are larger than $t$ is assumed to follow the power law 

\begin{equation}
P_t(\Theta)\propto \frac{1}{t}\left(\frac{\Theta}{t}\right)^{-\tau_b^{all}} H\left(\Theta-t\right) \:,
\label{eq:roux1}
\end{equation}
where $H(x)=1$ for $x>0$ and $H(x)=0$ otherwise. The dependency on $t$ in the prefactor follows from the normalization in the interval $t<\Theta<\infty$ \cite{roux1989}. We remind that in the IP model, the time corresponds to the size (mass) of the invaded cluster, measured in number of pores. For a cluster of size $\Theta$, the distribution $Q_\Theta(r)$ of distances $r$ between pores in that cluster is also assumed to follow a power law,
\begin{equation}
Q_\Theta(r)\propto \Theta^{-\left(1+\alpha\right)/D} \: r^\alpha H\left(\Theta^{1/D}-r\right) \:,
\label{eq:roux2}
\end{equation}
for $r$ in the interval $0<r<\Theta^{1/D}$, where again the dependency on $\Theta$ in the prefactor is obtained by the normalization in the interval $0<r<\Theta^{1/D}$ \cite{roux1989}. The pair correlation function $N(r,t)$ is then given by
\begin{equation}
N(r,t)=\int_0^\infty P_t(\Theta) Q_\Theta(r) d\Theta \:,
\label{eq:roux3}
\end{equation}
which, using Eqs.~(\ref{eq:roux1}) and (\ref{eq:roux2}) leads to

\begin{equation}
N(r,t) r = f\left(\frac{r^D}{t}\right) = \begin{cases}
\left(\frac{r^D}{t}\right)^\frac{1+\alpha}{D} &\text{if $\frac{r^D}{t} \ll 1$}\\
\left(\frac{r^D}{t}\right)^{1-\tau_b^{all}} &\text{if $\frac{r^D}{t} \gg 1$}
\label{eq:roux4}
\end{cases} \:.
\end{equation}

The insightful argumentation provided by Roux and Guyon \cite{roux1989} naturally generates the correct functional form of Eqs.~(\ref{eq:hat1}) and (\ref{eq:hat2}). Nevertheless their analysis for the numerical values of the exponents presents some inconsistent assumptions, as noted by Maslov \cite{maslov1995}. The exponent $\tau_b^{all}$ characterizing the distribution of all backwards avalanches was shown \cite{maslov1995} to be given by $\tau_b^{all}=3-\tau$, where $\tau=1+\left(D_e-1/\nu\right)/D$ is the cluster size distribution exponent (first derived in \cite{martys1991}, and verified experimentally in \cite{moura2017}), $D$ and $D_e$ are respectively the fractal dimensions of the growing cluster and its boundary and $\nu=4/3$ is the exponent characterizing the divergence of the correlation length \cite{feder1988,stauffer1994}. Using the values  $D=1.82$ and $D_e=4/3$ (external perimeter growth) \cite{feder1988,stauffer1994}, we find that the short-time exponent is

\begin{equation}
1-\tau_b^{all}=-1+\left(\frac{D_e-1/\nu}{D}\right) \implies 1-\tau_b^{all}=-0.68 \:,
\label{eq:shorttime}
\end{equation}
which is consistent with our measurements and very close to the value $-0.6$ reported in Ref.~\cite{furuberg1988}, see Fig.~\ref{fig:hat2}.

The computation of the long-time exponent brings additional challenges due to the difficulty in estimating the exponent $\alpha$ in Eq.~(\ref{eq:roux2}) \cite{roux1989}. With that in mind, we take an alternative approach here. Consider the situation in which a pore-throat that gives access to a pore at position $\bm{r_1}$ has been reached by the liquid air interface at a time $t_*$. From that moment on, that pore-throat and the pore-body at $\bm{r_1}$ are available to the invasion. Let us focus here on the case in which this particular pore-throat has a relatively high value of capillary pressure threshold, such that its invasion will typically take a long time to occur. This type of event lies in the long-time regime, for which $t \gg r^{D}$, and we are interested in addressing their relative probability of occurrence.

For the invasion of the pore at $\bm{r_1}$ to happen exactly at time $t=t_1$, a set of two well defined conditions must be verified at that time: 1) the capillary pressure must reach a historically high value since time $t=t_*$ and 2) the pore-throat at $\bm{r_1}$ must have the lowest value of capillary pressure threshold (i.e., be the widest pore-throat) among those that belong to the liquid air interface. Let us initially address condition 1). Consider the capillary pressure signal as the discrete time series formed by the sequence of the capillary pressure thresholds $p\left(t\right)$ associated with the invasion of successive pore-throats. If the system is in a statistical steady state (say, the flow has been going for a long time and happens in a long rectangular cell of large but finite width), $p\left(t\right)$ fluctuates around some well defined average and the historical maximum is equally likely to occur anywhere in the interval $t_*<t\leq t_1$. The probability that it occurs at the extreme point $t=t_1$ is then proportional to $1/\left(t_1-t_*\right)$. Once the capillary pressure reaches the historical maximum at $t=t_1$, the invasion at $\bm{r_1}$ \emph{can} happen. Next, we consider condition 2) and calculate the probability that the particular pore-throat in question is the one with the lowest threshold. Since the capillary pressure at time $t=t_1$ has reached a historical maximum, it is the first time that the pore-throats on the interface have been tested against such a high value of capillary pressure, and at this pressure they all have the same invasion probability. The number of sites $N_I$ that belong to the interface of the cluster that has grown since time $t=t_0$ (when the reference pore at $\bm{r_0}$ was invaded) scales as $N_I \propto t^{D_e/D}$. The probability that the particular pore-throat at $\bm{r_1}$ is the widest is simply given by $1/N_I$. (We have made the assumption that the invasion of the pore at $\bm{r_1}$ does not depend on invasion events that happened earlier than the invasion at $\bm{r_0}$. In the limit $t=t_1-t_0 \to \infty$, this approximation becomes exact). By considering the probability of simultaneously satisfying conditions 1) and 2), we have

\begin{equation}
N(r,t) \propto \frac{1}{\left(t_1-t_*\right)} \frac{1}{t^{D_e/D}} \approx \frac{1}{t^{\xi}} \:,
\label{eq:longtime1}
\end{equation}
where $\xi=1+D_e/D$ and, in the limit of large times, $t_1-t_* \approx t_1-t_0 = t$. The $r$ dependence in the previous equation is obtained by considering again Eq.~(\ref{eq:hat1}). Since the product $N(r,t) r$ is a function only of the reduced variable $r^D/t$, we have that in the long-time regime,

\begin{equation}
N(r,t) \propto \frac{1}{r} \left(\frac{r^D}{t}\right)^{\xi} \:.
\label{eq:longtime2}
\end{equation}
Using the literature values $D = 1.82$ and $D_e = 4/3$ \cite{feder1988,stauffer1994}, we find $\xi = 1.73$. This value is higher than the exponent $\xi = 1.4$ reported by Furuberg et al. \cite{furuberg1988}, but is consistent with the experimental data, particularly in the extreme left region of Fig.~\ref{fig:hat2}, where $t \gg r^{D}$ and the approximations of the model should hold best.

After the aged (hard) site at $\bm{r_1}$ is invaded, we could in principle have the invasion of easier pores in its vicinity that would also contribute to the counting in the long-time regime of $N(r,t)$. Although the argument made here counts explicitly only the aged (hard) sites, the invasion of easier ones in the long-time regime is conditioned to the prior invasion of an aged site and therefore should not change the temporal scaling of Eq.~(\ref{eq:longtime1}).

In this work we have given experimental validation to the unusual dynamic scaling for the pair correlation function $N(r,t)$ during the slow drainage of a porous medium, first observed by Furuberg et al. \cite{furuberg1988} nearly 30 years ago. Although this important result has been reproduced in other numerical works \cite{ferer2002}, to the best of our knowledge, the experimental verification presented here is new. By linking two important contributions to the literature, the works of Roux and Guyon \cite{roux1989} and Maslov \cite{maslov1995}, we found out that both approaches lead to the same predictions for the short-time exponent of $N(r,t)$, which agrees well with our measurements. A new theoretical explanation for the long-time exponent has also been provided, with good agreement with the experimental data. Possible extensions of the work include other flow regimes and geometries (e.g. radial injection). While $N(r,t)$ probably varies for faster flows, it possibly remains unchanged for slow drainage in a 2D radial geometry, once transient effects from the point inlet are dissipated.
\\

We gratefully acknowledge the support from the University of Oslo, University of Strasbourg, the Research Council of Norway through its Centre of Excellence funding scheme with project number 262644, the CNRS-INSU ALEAS program and the EU Marie Curie ITN FLOWTRANS network.

\bibliographystyle{ieeetr}
\bibliography{bib}

\end{document}